\documentclass{jaa}
\usepackage{graphicx}
\usepackage{pdflscape}
\usepackage{multirow}
\usepackage{longtable}
\begin{document}

%%paper title
%%For line breaks \\ can be used within title
\title{The Ratio of the Core  to the Extended Emissions \\in the Comoving Frame for Blazars}

%%author names are separated by comma (,)
%%use \and before the last author name
%%use a * along with the number separated by comma
%% for the  author for correspondence
%%\textsuperscript{number} is used for affiliation
%%\affilOne, \affilTwo etc., upto \affilTwentyfive is possible
%%Please note the first letter after \affil is capitalised in the command
%%

\author{Yun-Tian Li\textsuperscript{1,2}, Shao-Yu Fu\textsuperscript{2}, Huan-Jian Feng\textsuperscript{2}, Si-Le He\textsuperscript{2}, Chao Lin\textsuperscript{1,3}, Jun-Hui Fan\textsuperscript{1,3,*},  Denise Costantin\textsuperscript{1,3} \and Yu-Tao Zhang\textsuperscript{1,3}}
\affilOne{\textsuperscript{1}Center for Astrophysics, Guangzhou University,  Guangzhou 510006, China.\\}
\affilTwo{\textsuperscript{2}Dept. of Phys. School for Physics and Electronic Engineering, Guangzhou University, Guangzhou 510006, China.\\}
\affilThree{\textsuperscript{3}Astron. Sci. and Tech. Research Lab. of Dept. of Edu. of Guangdong Province, China.}

%%escape two column mode for title, affiliation and abstract
%%by giving \twocolumn command as shown

\twocolumn[{

\maketitle

%%include \corres to print the corresponding author Email id
\corres{fjh@gzhu.edu.cn}

%%include \msinfo for
%%manuscript information such as
%%received, revised and accepted dates
%%
\msinfo{}{}{}

%%abstract
\begin{abstract}
In a two-component jet model, the  emissions are the sum of the core and extended emissions: $S^{\rm ob}=S_{\rm core}^{\rm ob}+S_{\rm ext}^{\rm ob}$, with the core emissions, $S_{\rm core}^{\rm ob}= f S_{\rm ext}^{\rm ob}\delta^{q}$, being a function of the Doppler factor, $\delta$, the extended emission, $S_{\rm ext}^{\rm ob}$, jet type dependent factor, $q$, and the ratio of the core to the extended emissions in the comoving frame, $f$. The $f$ is an unobservable but important parameter. Following our previous work, we collect 65 blazars with available Doppler factor, $\delta$, superluminal velocity, $\beta_{app}$,  and core-dominance parameter, $R$, calculate the ratio, $f$, and peform statistical analyses. We find that the ratio, $f$, in BL Lacs is on average larger than that in FSRQs. We suggest that the difference of the ratio $f$ between FSRQs and BL Lacs is one of the possible reasons that cause the difference of other observed properties between them. We also find some significant correlations between $\log f$ and other parameters, including intrinsic (de-beamed) peak frequency, $\log \nu _{\rm p}^{\rm in}$, intrinsic polarization, $\log P^{\rm in}$, and core-dominance parameter, $\log R$,  for the whole sample. In addition, we show that the ratio, $f$, can be estimated by $R$.
\end{abstract}

%%insert keywords separated by 3 hyphens using \keywords{words}

}]
%%close the twocolumn escape here

%%include \doinum{number}for the DOI number in the header
%%include \volnum{number} for the volume number in the header
%%include \year{yyyy} for  year of publication in the header
%%include \pgrange{num--num} page range of article in the header
%%include \artcitid{num} for the article citation id
%%include \lp to print last page of the article
%%include \setcounter{page}{pagenum} for the exact starting page of the article

\doinum{12.3456/s78910-011-012-3}
\artcitid{\#\#\#\#}
\volnum{123}
\year{2016}
\pgrange{23--25}
\setcounter{page}{23}
\lp{25}

\section{Introduction}
Blazars are a particular subclass of the radio-loud active galactic nuclei (AGNs), which have high and variable polarization, large and rapid variation, high energetic $\gamma$-ray emissions, and superluminal motions, etc. Blazars can be divided into two subclassses, namely, flat spectrum radio quasars (FSRQs) and BL Lacertae objects (BL Lacs). The main difference between these two subclasses is that BL Lacs show no (or very weak) emission line features while FSRQs display strong emission lines. However, BL Lacs and FSRQs are quite similar in their continuum emission properties
(Fan 2003;
Fan \& Lin 2003;
Fan {\em et al.} 2009, 2014;
Gupta {\em et al.} 2009, 2016;
Lin \& Fan 2016;
Pei {\em et al.} 2016;
Lin {\em et al.} 2017
).

Following Abdo {\em et al.} (2010), blazars can be divided into
low synchrotron peaked (LSP, $\nu_{\rm{p}}^{s} <10^{14}$ Hz),
 intermediate synchrotron peaked  (ISP, $10^{14}$ Hz $<\nu_{\rm{p}}^{s} < 10^{15}$ Hz), and
 high synchrotron peaked (HSP, $\nu_{\rm{p}}^{s} > 10^{15}$ Hz) sources, from their spectral energy distributions (SEDs) (also see
  Ackermann {\em et al.} 2015;
  Fan {\em et al.} 2015).
In our recent work (Fan {\em et al.} 2016), the SEDs of 1392 blazars are fitted and the corresponding ${\log \nu^{s}_{p}}$ are obtained, and we classified the 999 Fermi blazars with available redshifts into
  LSP ($\nu_{\rm{p}}^{s} <10^{14}$ Hz),
  ISP ($10^{14}$ Hz $<\nu_{\rm{p}}^{s} < 10^{15.3}$ Hz),
  HSP ($\nu_{\rm{p}}^{s} > 10^{15.3}$ Hz) .

The extreme observational properties of blazars are attributed to a beaming effect. In a beaming model, the emissions can be separated into the jet and extended components in the comoving frame: $S^{\rm in}=S^{\rm in}_{j}+S^{\rm in}_{ext}=(f+1)S^{\rm in}_{ext}$, where $f=S^{\rm in}_{j}/S^{\rm in}_{ext}$ is the ratio of the intrinsic (de-beamed) jet to the extended emissions. Since the jet emissions are beamed in the observed frame $S^{\rm ob}_{j}=S^{\rm in}_{j} \delta^{q}$, the observed emissions can be expressed in a form: $S^{\rm ob}=(f \delta^{q}+1)S^{\rm ob}_{ext}$, where $\delta=[\Gamma(1-\beta\cos\theta)]^{-1}$ is a Doppler factor, $q$ is a jet type depended parameter: $q = 2 + \alpha$ for continuous jet, $q = 3 + \alpha$ for a jet with distinct ``blobs'' (Lind \& Blandford 1985), $\alpha$ is a spectral index, $S_{\nu} \propto \nu^{-\alpha}$, $\theta$ is the viewing angle, $\beta $ is the jet speed in unit of the speed of light, and $\Gamma =1/(1-\beta^2)^{1/2}$ is the bulk Lorentz factor. In a two-component jet model, the beamed emissions correspond to the core emissions, $S_{\rm core}$, while the unbeamed ones are linked to the extended emissions, $S_{\rm ext}$. Then a core-dominance parameter, $R$, can be defined as \begin{equation}
R=S^{\rm ob}_{\rm core}/S^{\rm ob}_{\rm ext}=f\delta^{q}.
\end{equation}
When the viewing angle, $\theta$, is large, the emission from the receding jet is no longer negligible, then Eq. (1) is replaced by (Orr \& Brown 1982, see also Urry \& Padovani 1995)
\begin{equation}
R(\theta)=f\{[\Gamma(1-\beta cos\theta)]^{-q}+[\Gamma(1+\beta cos\theta)]^{-q}\}
\end{equation}
At a viewing angle of 90 degree, the relation between ratio $f$ and $R_T$ is given by
\begin{equation}
R_T = R(90^\circ) = 2f/\Gamma^{q}.
\end{equation}

Facing the differences and similarities between BL Lacs and FSRQs, some authors suggested that there is an evolution tendency between BL Lacs and FSRQs (e.g., Sambruna {\em et al.} 1996). However, no significant difference of black hole mass was found between them in Wu {\em et al.} (2002).
In 2003, we compiled 41 sources (10 BL Lacs, 27 FSRQs, 4 radio galaxies) with available superluminal velocity, Doppler factor from L\"ahteenim\"aki \& Valtaoja (1999, hereafter LV99) and core dominance parameter, calculated the ratio, $f$, found the $f$ values in BL Lacs are larger than that in FSRQs, and proposed that the difference in emission line feature between BL Lacs and FSRQs is from their difference in ratio, $f$ (Fan 2003).

Following our previous work (Fan 2003), we compile 65 blazars (with 28 new sources) in order to calculate the ratio, $f$, and then do some statistical analysis.
This work is arranged as follows: we will describe the sample and the results in section 2, and give some discussions and conclusions in section 3.

\section{Sample and Results}

From Eq. (2), we can obtain the ratio (Fan 2003)
\begin{equation}
f=R(\theta)/\{[\Gamma(1-\beta cos\theta)]^{-q}+[\Gamma(1+\beta cos\theta)]^{-q}\}
\end{equation}
Although $\Gamma$ and $\theta$ are unobservable parameters, they can be obtained if the Doppler factor, $\delta$, and apparent velocity, $\beta_{app}$ are known, as
\begin{equation}
\Gamma=(\beta_{app}^2+\delta ^2+1)/2\delta,
\end{equation}
\begin{equation}
\tan \theta=2\beta_{app}/(\beta_{app}^2+\delta^2-1),
\end{equation}
Thus, from Eq. (4), $f$ can be obtained for a superluminal jet with available apparent velocity, $\beta_{app}$, Doppler factor, $\delta$, and core-dominance parameter, $R(\theta)$.

In a two-component model,  we assumed that the polarized emissions are only from the jet and the extended emissions are not polarized. In this case, we assumed that the polarized and unpolarized emissions are proportional to each other (Fan {\em et al.} 1997), namely $S_{j} = S_{j p} + S_{j up} = (1+\eta)S_{j up}$, where $\eta = S_{j p} / S_{j up}$. Then,
the frequency and polarization at the observers' frame can be expressed as  (e.g., Fan {\em et al.} 1997, 2006; Fan 2002),
\begin{equation}
\nu^{\rm ob} =\delta \nu^{\rm in}/ (1+z)
\end{equation}
\begin{equation}
P^{\rm ob} = \frac{(1+f)\delta^q}{1+f\delta^q} P^{\rm in}
\end{equation}
where $z$ is a redshift, $\nu^{\rm in}$ is an intrinsic frequency, $P^{\rm in}$ is an intrinsic total polarization, which defined as
\begin{equation}
P^{\rm in}=\frac{f}{1+f}P_{j}^{\rm in}=\frac{f}{1+f}\frac{\eta}{1+\eta}
\end{equation}
where $P_{j}^{\rm in}$ is the intrinsic jet polarization. Therefore, when $f$ is obtained and $P^{\rm ob}$ is known, $P^{\rm in}$ can be obtained by Eq. (8), and then $\eta$ can be obtained from $P^{\rm in}$ and $f$.

\subsection{Sample}
In this work, we compile a sample of 65 blazars (18 BL Lacs, 47 FSRQs). They are listed in Table 1, in which,
Col. (1) gives IAU name,
Col. (2) other name,
Col. (3) redshift,
Col. (4) classification, ``Q" stands for FSRQs and ``B" for BL Lacs,
Col. (5) apparent speed in units of speed of light,
Col. (6) Doppler boosting factor,
Col. (7) references of Col. (5) and (6),
Cols. (8) and (9) core-dominance parameter and the corresponding references,
Col. (10) synchrotron peak frequency ${\log \nu^{\rm s}_{\rm p}}$ from Fan {\em et al.} (2016).
For the sources with no available ${\log \nu^{s}_{p}}$ in Fan {\em et al.} (2016), we use the empirical relationship introduced in Fan {\em et al.} (2016) to estimate it, as follow
\begin{equation}
{\rm log}\nu_p^s=
  \begin{cases}
    \,\, 16 + 4.238X  \,\,\,\,\,\,\,X<0  \\
    \,\, 16 + 4.005Y  \,\,\,\,\,\,\, X>0,
  \end{cases}
\end{equation}
where $X = 1 - 1.262 \alpha_{\rm RO} - 0.623 \alpha_{\rm OX} $, and $Y = 1.0 + 0.034 \alpha_{\rm RO} - 0.978 \alpha_{\rm OX}$, $\alpha_{\rm RO}$ and $\alpha_{\rm OX}$ are the radio-to-optical and optical-to-X-ray effective spectral indexes.
The corresponding effective spectral indexes are from BZCAT catalog (Massaro {\em et al.} 2015),
they are labeled with a ``$\ast$",
Cols. (11) and (12) give the optical polarization and the corresponding references.

\subsection{Result}

\subsubsection{Averaged Result}

Based on the data listed in Table 1, we calculate the Lorentz factor, $\Gamma$, the viewing angle, $\theta$, the jet speed, $\beta$, the ratio, $f$, the core-dominance parameter at 90 degree, $R_T$, the intrinsic peak frequency, $\nu_{\rm p}^{\rm in}$, the intrinsic polarization, $P^{\rm in}$, and the ratio, $\eta$ for the 65  blazars.
Those calculated values are listed in Table 2. The corresponding range and averaged values of $\delta$, $\beta_{app}$, $\log R$, $\log \nu _{\rm p}^{\rm s}$, $\log P^{\rm ob}$, $\Gamma$, $\theta$, $\beta$, $\log f$, $\log R_T$, $\log\nu_{\rm p}^{\rm in}$ $\log P^{\rm in}$ and $\eta$ are listed in Table \ref{tbl:Averaged} for the whole sample, and for BL Lacs and FSRQs separately.

\begin{figure}[tb]
  \centering
  \includegraphics[width=\columnwidth]{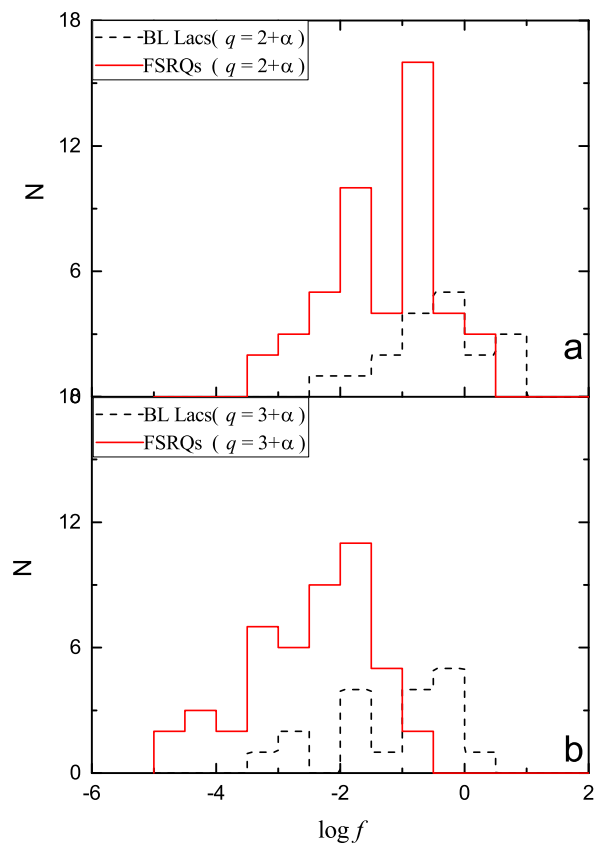}\\
  \caption{The distribution of the ratio, $\log f$, for the case of $q=2+\alpha$ (a) and $q=3+\alpha$ (b), solid lines are FSRQs and broken {\bf lines are} BL Lacs.}
  \label{Lin-2016-lgf}
\end{figure}

From Table \ref{tbl:Averaged}, we have that $\log f$ is in a range of $-3.11$ to 0.98, with a averaged value of $<\log f>=-1.06\pm 0.93$ for the case of $q=2+\alpha$, and in a range of $-4.65$ to 0.22, with $<\log f>=-2.09\pm 1.15$ for $q=3+\alpha$ for the whole sample.
When the subclasses are considered, we have $<\log f>=-0.40\pm0.88$ $(q=2+\alpha)$ and $<\log f>=-1.17\pm0.99$ $(q=3+\alpha)$ for BL Lacs, $<\log f>= -1.31\pm0.84$ $(q=2+\alpha)$ and $<\log f>=-2.45\pm1.01$ $(q=3+\alpha)$ for FSRQs.
One sample Kolmogorov-Smirnov (K-S) test indicates that the ratio, $f$, follows a lognormal distribution with significant levels being $p=99.7\%$ ($q=2+\alpha$) and $p=55.5\%$ ($q=3+\alpha$) for BL Lacs; $p=25.6\%$ ($q=2+\alpha$) and $p=40.3\%$ ($q=3+\alpha$) for FSRQs.
Moreover, for the ratio, $\log f$, K-S test gives that the probability for BL Lacs and FSRQs distributions to be from the same distribution is $1.86\times10^{-3}$ for $q=2+\alpha$ and $2.20\times10^{-4}$ for $q=3+\alpha$.
%In addition, Levene's test (F-test) shown the equality of variances is in significant level $44.47\%$ for $p=2+\alpha$, and $60.55\%$ for $p=3+\alpha$, thus the variances of FSRQs and BL Lacs are mainly equivalent.
From a T-test, the probability that BL Lacs and FSRQs have the same averaged value of $\log f$ is $2.87\times10^{-4}$ for $q=2+\alpha$, and $2.20\times10^{-5}$ for $q=3+\alpha$, and the averaged difference in $\log f$ between them is $\Delta (\log f) = 0.90 \pm 0.24$ for $q=2+\alpha$, and $\Delta (\log f) = 1.28 \pm 0.28$ for $q=3+\alpha$. Therefore, the distribution and the averaged value of $\log f$ in BL Lacs are different from those in FSRQs, see also Fig. \ref{Lin-2016-lgf}.

Some researches (eg., Abdo {\em et al.} 2010, Fan {\em et al.} 2016) found that peak frequency in BL Lacs are different with that in FSRQs.
In this work, we investigate the difference between them in $\log\nu_{\rm p}^{\rm in}$ parameter, and find that $\log\nu_{\rm p}^{\rm in}$ is in a range of 11.07 to 14.62, with an averaged value of $<\log\nu_{\rm p}^{\rm in}>=12.69\pm 0.77$ for the whole sample. Considering the subclasses, we have that $\log\nu_{\rm p}^{\rm s}$ is 12.04 to 14.62 with $<\log\nu_{\rm p}^{\rm in}>=13.11\pm 0.83$ for BL Lacs, $\log\nu_{\rm p}^{\rm in}$ is 11.07 to 13.97 with $<\log\nu_{\rm p}^{\rm in}>=12.52\pm 0.69$ for FSRQs. And the probability (K-S test) for BL Lacs and FSRQs distributions to be from the same distribution is $3.62\%$ in $\log \nu_{\rm p}^{\rm in}$.
Thus, BL Lacs show not only larger ratio, $f$, but also higher intrinsic peak frequency than FSRQs.

\subsubsection{Correlations}
\begin{figure}[t]
  \centering
  \includegraphics[width=\columnwidth]{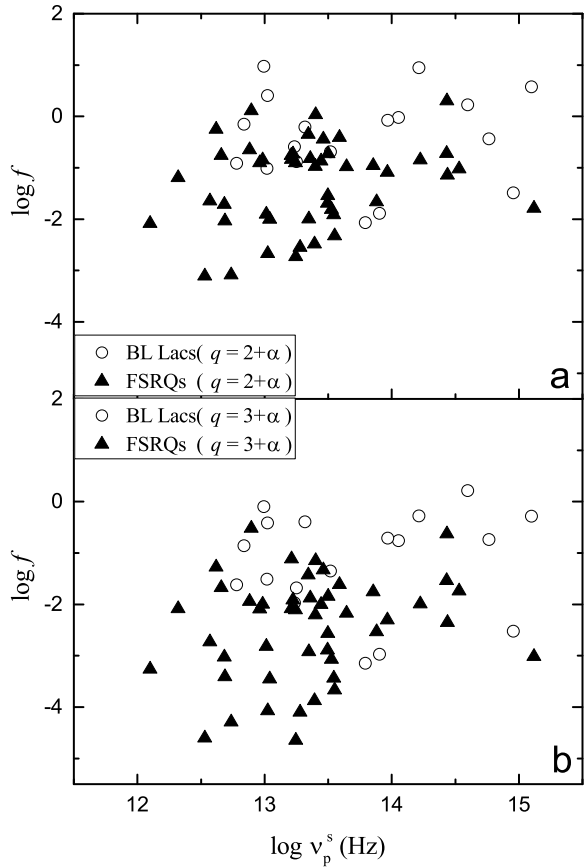}\\
  \caption{Plot of the ratio, $f$, against the peak frequency for the case of $q=2+\alpha$ (a) and $q=3+\alpha$ (b), the circle symbols stand for BL Lacs and the triangle symbol stand for FSRQs.}
  \label{Lin-2016-logf-lognu}
\end{figure}

\begin{figure}[h]
  \centering
  \includegraphics[width=\columnwidth]{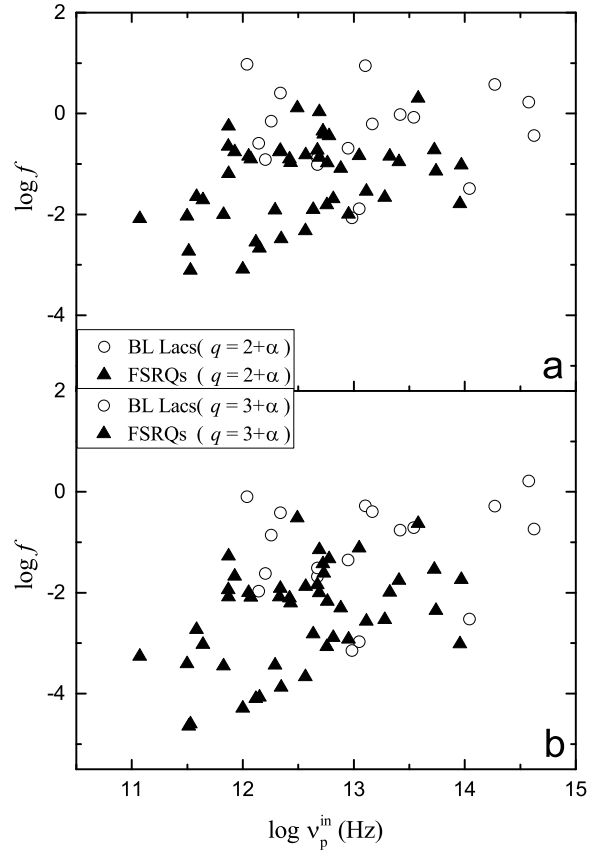}\\
  \caption{Plot of the ratio, $f$, against the intrinsic peak frequency for the case of $q=2+\alpha$ (a) and $q=3+\alpha$ (b), the circle symbols stand for BL Lacs and the triangle symbol stand for FSRQs.}
  \label{Lin-2016-logf-lognuin}
\end{figure}

\begin{figure}[t]
  \centering
  \includegraphics[width=\columnwidth]{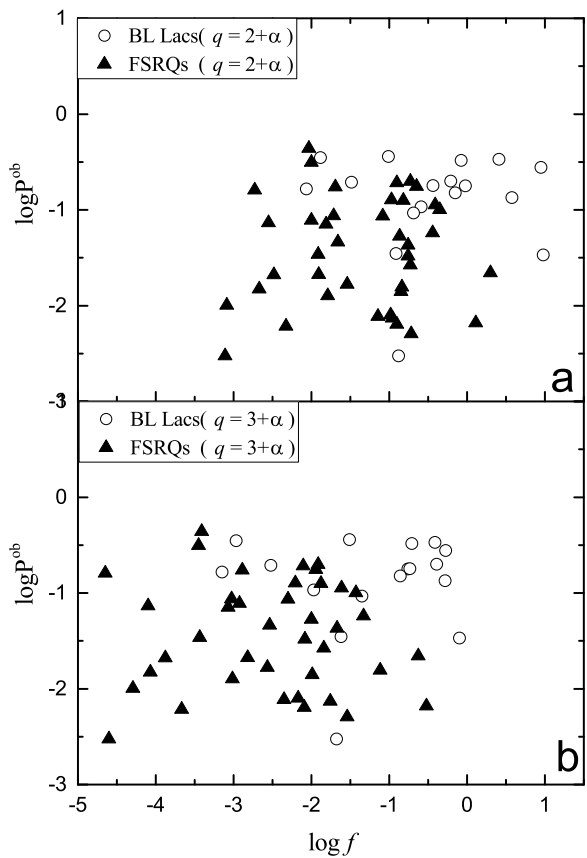}\\
  \caption{Plot of the observed polarization against the ratio, $f$, for the case of $q=2+\alpha$ (a) and $q=3+\alpha$ (b), the circle symbols stand for BL Lacs and the triangle symbol stand for FSRQs.}
  \label{Lin-2016-logP-logf}
\end{figure}

\begin{figure}[h]
  \centering
  \includegraphics[width=\columnwidth]{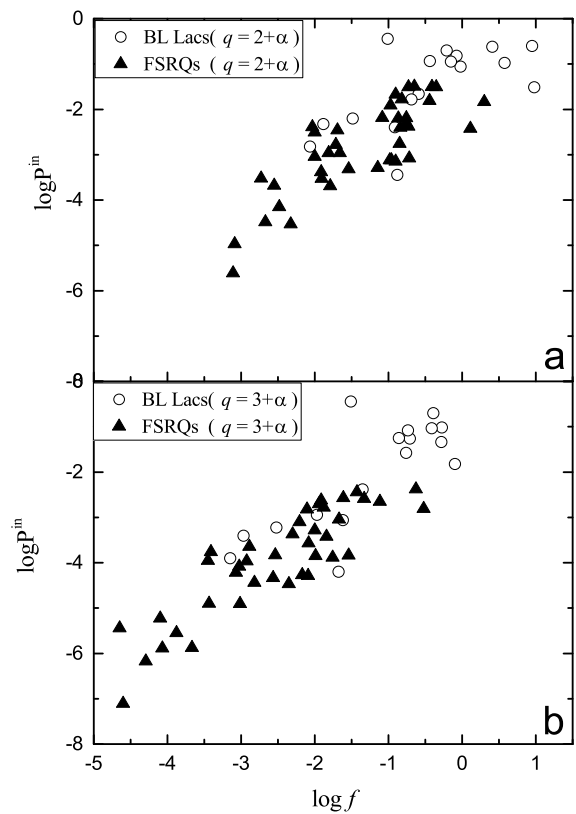}\\
  \caption{Plot of the intrinsic polarization against the ratio, $f$, for the case of $q=2+\alpha$ (a) and $q=3+\alpha$ (b), the circle symbols stand for BL Lacs and the triangle symbol stand for FSRQs.}
  \label{Lin-2016-logPin-logf}
\end{figure}

\begin{figure}[t]
  \centering
  \includegraphics[width=\columnwidth]{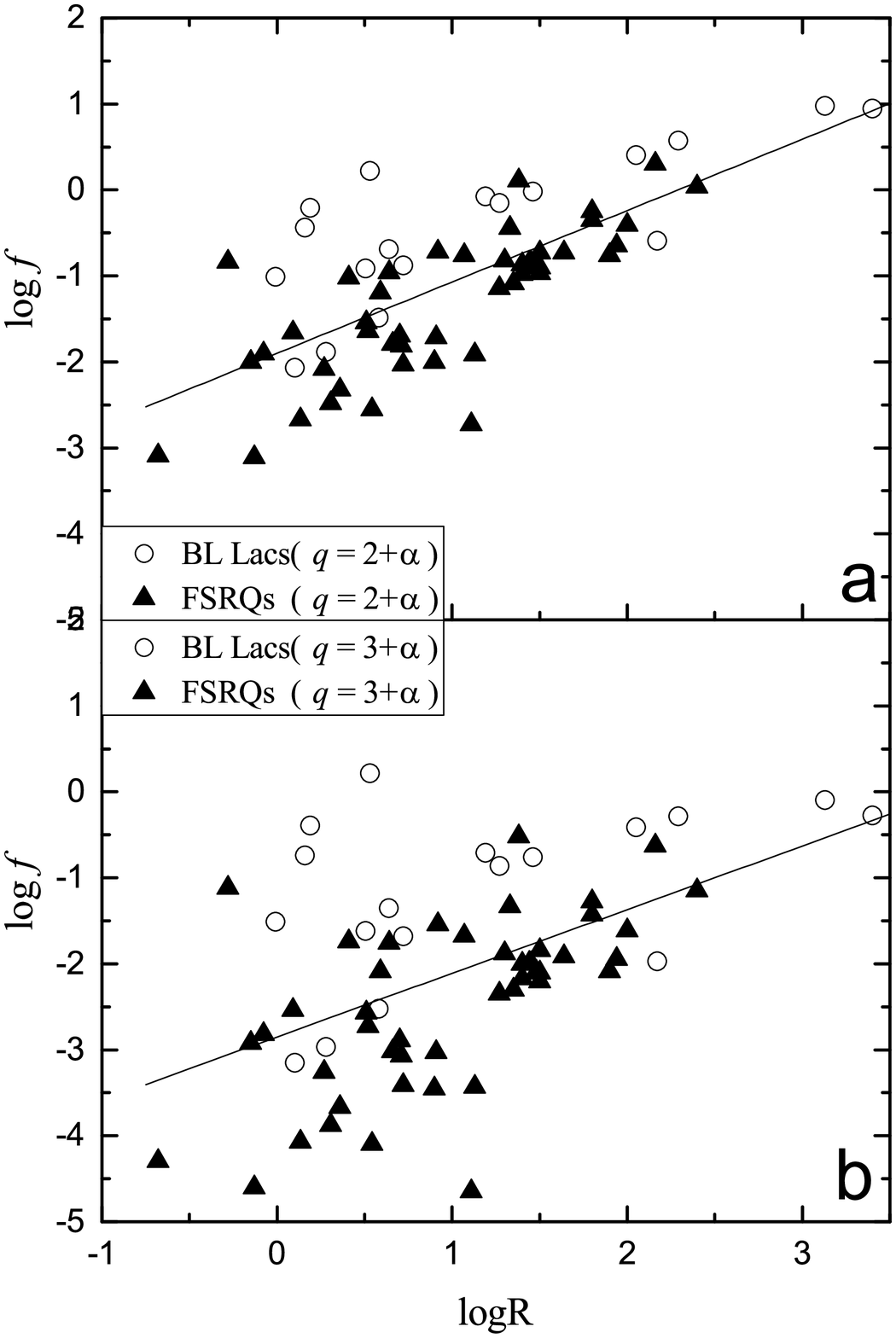}\\
  \caption{Plot of the core-dominance parameter against the ratio, $f$, for the case of $q=2+\alpha$ (a) and $q=3+\alpha$ (b), the circle symbols stand for BL Lacs and the triangle symbol stand for FSRQs.}
  \label{Lin-2016-logf-logR}
\end{figure}

 We investigate the correlations between $\log f$ and several parameters (including $\log \nu^{\rm s}_{\rm p}$, $\log \nu _{\rm p}^{\rm in}$, $\log P^{\rm ob}$, $\log P^{\rm in}$, and $\log R$) for the whole sample, as well as the subclasses. The linear regression method is adopted to the correlation analysis, then the corresponding results are listed in Table \ref{tbl:Correlation} and shown in Figs. \ref{Lin-2016-logf-lognu}--\ref{Lin-2016-logf-logR}.

\underline{$f$ vs $\nu_{\rm p}^{\rm s}$}: Correlations are found between $\log f$ and $\log \nu _{\rm p}^{\rm s}$ for the whole sample with a correlation coefficient $r=0.26$ and a chance probability $p=3.63\%$ for the case of $q=2+\alpha$, see Fig. \ref{Lin-2016-logf-lognu}. There are also correlations between $\log f$ and $\log \nu _{\rm p}^{\rm in}$ with $r=0.37$  ($p=2.50 \times 10^{-3}$) for the whole sample, and $r=0.37$ ($p=1.13\%$) for FSRQs for $q=2+\alpha$, but no correlation is found for BL Lacs with $r=0.03$ ($p=90.01\%$) for $q=2+\alpha$, see Fig. \ref{Lin-2016-logf-lognuin}. Some similar results are found in the case of $q=3+\alpha$, see Table \ref{tbl:Correlation}.

\underline{$f$ vs $P$}: The correlations between $\log f$ and $\log P^{\rm ob}$ and between $\log f$ and $\log P^{\rm in}$ are also investigated. No correlation is found between $\log f $ and $\log P^{\rm ob}$ for the whole sample, BL Lacs or FSRQs respectively with chance probabilities being 10.02\%, 97.35\% and 73.01\%. However, strong correlations are found between $\log f$ and $\log P^{\rm in}$ for the whole sample, BL Lacs, and FSRQs, with $r=0.84$ ($p=6.21\times 10^{-16}$), $0.74$ ($7.55\times 10^{-4}$), and $0.82$ ($1.48\times 10^{-10}$) for $q=2+\alpha$ respectively. The corresponding figures are shown in Figs. \ref{Lin-2016-logP-logf} and \ref{Lin-2016-logPin-logf}. Even stronger correlations (see Fig. \ref{Lin-2016-logPin-logf} and Table \ref{tbl:Correlation}) are seen for $p=3+\alpha$.

\underline{$f$ vs $R$}: We find significant correlations between $\log f$ and $\log R$ with
$ r = 0.72$ ($ p=1.57 \times 10^{-11}$) for the whole sample,
$r=0.77$ ($p=2.19\times 10^{-4}$) for BL Lacs,
and $r=0.77$ ($p=3.13\times 10^{-10}$) for FSRQs for $q=2+\alpha$.
Some similar results are found in the case of $q=3+\alpha$, see Table \ref{tbl:Correlation}.
The corresponding figures are shown in Fig. \ref{Lin-2016-logf-logR}. Those results indicate that the ratio, $f$, is directly correlated with a core-dominance parameter ($f \sim R^{0.8}$) regardless of different beaming boosting factors and viewing angles of different sources. The significant correlation between $\log f$ and $\log R$ indicates that the unobservable ratio, $f$, can be estimated from the core-dominance parameter, $R$, namely $\log f = (0.83\pm0.10) \log R -(1.90\pm0.13)$ for $q=2+\alpha$ and $\log f = (0.74\pm0.15) \log R -(2.85\pm0.20)$ for $q=3+\alpha$, see Table \ref{tbl:Correlation} and Fig. \ref{Lin-2016-logf-logR}.

\section{Discussions and Conclusions}
A relativistic beaming model is proposed to explain extreme observational properties for blazars.
Some authors (eg., Blandford \& K\"onigl 1979; Antonucci \& Ulvestad 1985) proposed that the difference of emission line feature between FSRQs and BL Lacs is due to beaming effect. As discussed in our previous work (Fan 2003), if the weak emission line means that the Doppler-boosted emissions dominate the isotropic line emissions, BL Lacs, which have weak emission lines, should have stronger beaming effect than FSRQs. However, the extreme observational properties of FSRQs indicate that the beaming effect of FSRQs is stronger than that of BL Lacs.
Others (eg., Padovani 1992, Urry \& Padovani 1995) proposed an evolution tendency between FSRQs and BL Lacs, with strong emission line FSRQs evolving into weak line BL Lacs. However there is no difference in black hole masses between FSRQs and BL Lacs in Wu {\em et al.} (2002) and our previous work (Fan 2005). So it is very difficult to explain the emission line properties of blazars in that way. Based on all those observational properties, we proposed that a good method should explain both the differences and similarities between FSRQs and BL Lacs, and the ratio, $f$, played an important role on the emission line feature of blazars (Fan 2003).

\subsection{Averaged Values}

In the present work, we have $<\log f>=-1.06\pm0.93$ ($q=2+\alpha$), and $<\log f>=-2.09\pm1.15$ ($q=3+\alpha$) for the whole sample.
Some authors suggest that the possible value of the ratio, $f$, is from 0.001 to 1.0
(Padovani \& Urry 1990, 1991;
 Urry {\em et al.} 1991;
 Urry \& Padovani 1995;
 Fan {\em et al.} 1997).
 Our present results ($\log f=-3.11$ to 0.98 for $q=2+\alpha$ and $\log f=-4.65$ to 0.22 for $q=3+\alpha$) are not in conflict with theirs ($\log f=-3$ to 0), but our sample shows a somewhat wider range.

In Fan (2003), we found $<\log f> = -0.99 \pm 0.22$ for the whole 41 blazars, $<\log f> = 0.11 \pm 0.49$ for BL Lacs, and $<\log f> = -1.59 \pm 0.19$ for FSRQs for the case of $q=3+\alpha$.
A K-S test shows that the probability for the $f$ distributions of BL Lacs and FSRQs to be from the same one is $3\times10^{-3}$. And the difference between them is $\Delta (\log f) = 1.68 \pm 0.52$. In this work, $\Delta (\log f) = 0.90 \pm 0.24$ for $q=2+\alpha$, and $\Delta (\log f) = 1.28 \pm 0.28$ for $q=3+\alpha$ are found from the T-test, and the K-S test shows clear difference of $\log f$ between BL Lacs and FSRQs in distributions. In addition, Fan (2002) found that $f_{\rm BL} \sim 15 f_{\rm FSRQs}$, namely $\Delta {\log f} = 1.17$. Thus our present results confirm others' and our previous ones.

In this work, we obtain $R_T$ for the whole sample. Since $R_T$ obey the lognormal distribution, but do not obey a normal distribution, we take the logarithm of it.
Then we find an averaged value $<\log R_T>=-2.91 \pm 1.30$ for $q=2+\alpha$ and $<\log R_T>=-5.02 \pm 1.79$ for $q=3+\alpha$ for the whole sample.
Orr \& Browne (1982) suggested to use $\Gamma=5$ and $R_T=0.0024$ to estimate the expected $R$ distribution. Their $R_{T}$ gives $\log R_{\rm T}=-2.62$. Our $<\log R_{\rm T}>=-2.91 \pm 1.30$ is accord with their $\log R_{\rm T}=-2.62$. Fan {\em et al.} (2004) got an averaged value of $\log R_T$, namely $<\log R_T>=-2.10 \pm 0.73$ (BL Lacs) and $<\log R_T>=-3.83 \pm 0.24$ (FSRQs) for the case of $q=3+\alpha$. Our results are $<\log R_T>=-3.39\pm1.72$ (BL Lacs) and $<\log R_T>=-5.65\pm1.39$ (FSRQs), which are not inconsistent with our previous results.

Some researches show that BL Lacs have, on average, higher observed polarization than do FSRQs (e.g., Fan {\em et al.} 2002; Yang {\em et al.} 2010).
In our sample, some similar results are found, namely $<\log P^{\rm ob}>=-0.90\pm 0.52$ (BL Lacs) and $<\log P^{\rm ob}>=-1.44\pm 0.56$ (FSRQs).
For intrinsic polarization, $<\log P^{\rm in}>=-1.58\pm 0.81$ (BL Lacs) and $<\log P^{\rm in}>=-2.84\pm 0.99$ (FSRQs) for $q=2+\alpha$, $<\log P^{\rm in}>=-2.19\pm 1.06$ (BL Lacs) and $<\log P^{\rm in}>=-3.95\pm 1.15$ (FSRQs) for $q=3+\alpha$ are found.
From a T-test, the probability that BL Lacs and FSRQs have the same averaged value of $\log P^{\rm in}$ are $2.50 \times 10^{-5}$ ($q=2+\alpha$) and $1.00 \times 10^{-6}$ ($q=3+\alpha$), which suggest that $\log P^{\rm in}$ of BL Lacs is higher than that of FSRQs.

Interestingly, the values of $\eta$ are almost the same for the two jet cases ($q=2+\alpha$ and $q=3+\alpha$). For both jet cases, we have $<{\eta}>=0.15\pm0.20$ for the whole sample, $<{\eta}> =0.29\pm0.24$ for BL Lacs, $<{\eta}> = 0.10 \pm0.15 $ for FSRQs, see also Table \ref{tbl:Averaged}. And the probability (T-test)
that BL Lacs and FSRQs have the same averaged value of $\eta$ is $4.33\times10^{-4}$, and the average difference is $\Delta ( {\eta} ) = 0.19 \pm 0.05$. The difference of $\eta$ between FSRQs and BL Lacs may be caused by differences in the magnetic field in their jets.

\subsection{Correlations}

For correlation analyses, there are correlations between $\log f$ and $\log \nu _{\rm p}^{\rm in}$ for the whole sample and FSRQs, but no correlation for BL Lacs, see Table \ref{tbl:Correlation} and Fig. \ref{Lin-2016-logf-lognu}. In our sample, $\log \nu _{\rm p}^{\rm in}$ is in a range of 11.07 to 14.62. In the AGNs model, FSRQs have smaller viewing angles and so superluminal motions are easier to detect than in BL Lacs (Urry \& Padovani 1995). Thus, the bias of our sample is possibly caused by the bias of detection of superluminal motion for blazars. Even though our sample is not complete in blazars, the one sample K-S test indicates that their subclasses follow normal distributions in $\log f$, indicating that they can represent a complete sample in the low peak frequency range. However, the correlation between $\log f$ and $\log \nu _{\rm p}^{\rm in}$ may be eliminated by the limited range of peak frequencies.
Thus, we need more blazars, especially HSP blazars, to investigate the correlation further.

No correlation is found between $\log f$ and $\log P^{\rm ob}$ for the whole sample ($r=0.22$, $p = 10.02\%$), BL Lacs ($r=0.01$, $p=97.35\%$) or FSRQs ($r=0.06$, $p=73.01\%$), see Fig. \ref{Lin-2016-logP-logf} and Table \ref{tbl:Correlation}.
Our calculation show that $f$ is much smaller than 1 for most sources. From Eq. (9), if $f$ is much smaller than 1 and $\eta$ is a constant for a specific group, then $\log P^{\rm in}$ can  be expressed in a form
\begin{equation}
\log P^{\rm in} \approx \log f + \log(\frac{\eta}{1+\eta}) \propto \log f
\end{equation}
Thus, a linear correlation should be expected between $\log P^{\rm in}$ and $\log f$ with a slope of 1.
That means, for a beaming model, if a sample belongs to a group, their $\log P^{\rm in}$ should be determined by $\log f$.
In the present work, strong positive correlations are found between $\log f$ and $\log P^{\rm in}$ for the whole sample ($r=0.84$, $p=6.21\times10^{-16}$), BL Lacs ($r=0.74$ $p=7.55\times10^{-4}$) and FSRQs ($r=0.82$, $1.48\times10^{-10}$) for $q=2+\alpha$, see Fig. \ref{Lin-2016-logPin-logf}. The slopes of those correlations are $0.96 \pm 0.09$ ($q=2+\alpha$), $1.07 \pm 0.07$ ($q=3+\alpha$) for the whole sample, $0.67 \pm 0.16$ ($q=2+\alpha$), $0.95 \pm 0.14$ ($q=3+\alpha$) for BL Lacs, and $0.95 \pm 0.11$ ($q=2+\alpha$), $0.97 \pm 0.09$ ($q=3+\alpha$) for FSRQs, which are mainly accord with the expected relation, especially for the whole sample and FSRQs.
The results indicate that the difference of ratio, $f$, alone cannot cause the difference in the observed polarization although it really affects the  intrinsic polarization for blazars.

For $f$ and $R$, we find strong positive correlations between $\log f$ and $\log R$ with $r=0.72$ ($p=1.57\times10^{-11}$) for $q=2+\alpha$, and $r=0.52$ ($p =8.00\times 10^{-6}$) for $q=3+\alpha$ for the whole sample, and their slopes are close to 0.8. Those results suggest that $\log f$ is correlated with $\log R$ regardless of different beaming boosting factors and viewing angles of different sources. We also find $\log f$ of BL Lacs is higher than that of FSRQs on average, which can be used to explain the difference of their core-dominance parameter.
What's more, a significant correlation between $\log f$ and $\log R$ is found in this work. Thus, we propose that the core-dominance parameter, $R$, can be used to estimate the unobservable parameter $f$, namely $\log f = (0.83\pm0.10) \log R -(1.90\pm0.13)$ for $q=2+\alpha$ and $\log f = (0.74\pm0.15) \log R -(2.85\pm0.20)$ for $q=3+\alpha$, see Fig. \ref{Lin-2016-logf-logR}.

\subsection{Conclusions}
In this work, we collect 65 blazars with available Doppler factor, $\delta$, superluminal velocity, $\beta_{app}$,  and core-dominance parameter, $R$. Then the ratio, $f$, of the comoving emissions in the jet to the extended emissions is calculated. We investigate the difference of the ratio, $f$, between BL Lacs and FSRQs. In addition, the correlations between $\log f$ and several parameters, including peak frequency, $\log \nu _{\rm p}$, polarization, $\log P$, and core-dominance parameter, $\log R$, are also studied. The corresponding results are listed in Tables \ref{tbl:Averaged} and \ref{tbl:Correlation}, and shown in Figs. \ref{Lin-2016-lgf}--\ref{Lin-2016-logf-logR}.
Our main conclusions are as follows.

1) The difference in averaged logarithm of ratio, $f$ between BL Lacs and FSRQs are $\Delta (\log f) = 0.90 \pm 0.24$ for the case of $q=2+\alpha$, and $\Delta (\log f) = 1.28 \pm 0.28$ for $q=3+\alpha$.
The difference of $\log f$ between FSRQs and BL Lacs is one of the possible reasons that cause the the difference in some observed properties between them, confirming our early result (Fan 2003);

2) There are clear correlations between $\log f$ and $\log \nu _{\rm p}^{\rm in}$ for the whole sample and FSRQs, but no correlation for BL Lacs;

3) No correlation is found between $\log f $ and $\log P^{\rm ob}$. Strong positive correlations are found between $\log f$ and $\log P^{\rm in}$ for the whole sample and the subclasses. The slopes of those correlations are consistent with an excepted relation. Those results indicate that the difference of the ratio causes the difference of intrinsic polarization but it can not alone cause the difference in observed  polarization  between BL Lacs and FSRQs;

4) Strong positive correlations between $\log f$ and $\log R$ are found, which suggests that the ratio, $f$ can be estimated by the core-dominance parameter, $R$.

\section*{Acknowledgement}
This work is supported by the National Natural Science Foundation of China (U1531245, U1431112, 11203007, 11403006, 10633010, 11173009, 11403006), and the Innovation Foundation of Guangzhou University (IFGZ),
Guangdong Innovation Team (2014KCXTD014),
Guangdong Province Universities and Colleges Pearl River Scholar
Funded Scheme (GDUPS) (2009), Yangcheng Scholar Funded
Scheme (10A027S), and supported for Astrophysics  Key Subjects of Guangdong Province and Guangzhou City.

%%use \balance somewhere in the left column of the last page to balance the two columns in the end page
\newpage
%%References section

%\begin{landscape}
\begin{table*}
\centering
\begin{scriptsize}
%\tabularfont
\caption{Blazars Sample}\label{tbl:Sample}
\begin{tabular}{llcccccccccc}
\topline
%\rotate
\textbf{Name}&\textbf{other name}&\textbf{$z$}&\textbf{Class}&\textbf{$\beta_{app}$}&\textbf{$\delta$}&\textbf{ref}&\textbf{$\log R$}&\textbf{ref}&\textbf{$\log \nu^{\rm s}_{\rm p}$}&\textbf{$P$(\%)}&\textbf{ref}\\
%\midline
    (1) & (2) & (3) & (4) & (5) & (6) & (7) & (8) & (9) & (10) & (11) & (12)\\
    \hline
    0003-066 & PKS 0003-066 & 0.347 & B     & 0.9   & 5.1   & S10   & 0.504  & F10   & 12.78$^{\ast}$  & 3.5   & X16 \\
    0016+731 & 1Jy 0016+731 & 1.781 & Q     & 6.7   & 7.8   & S10   & 0.59  & F11   & 12.32$^{\ast}$  &       &   \\
    0059+581 & TXS 0059+581 & 0.644 & Q     & 11.1  & 14.1  & H09   & 1.451  & F10   & 12.99$^{\ast}$  &       &      \\
    0106+013 & PKS 0106+01 & 2.099 & Q     & 26.5  & 18.2  & S10   & 0.71  & F11   & 13.53  & 7.1   & F02   \\
    0133+476 & S4 0133+47 & 0.859 & Q     & 13    & 20.5  & S10   & 0.91  & F11   & 12.69  & 8.7   & W11   \\
    0202+149 & 4C +15.05 & 0.405 & Q     & 6.4   & 15    & S10   & 0.27  & F11   & 12.10  &       &       \\
    0212+735 & S5 0212+73 & 2.367 & Q     & 7.6   & 8.4   & S10   & -0.15  & F11   & 13.35  & 7.8   & F02   \\
    0219+428 & 3C 66A    & 0.444 & B     & 14.89 & 1.99  & F04   & 0.16 & F11    & 14.76  & 18.0 & F04     \\      %++++++
    0224+671 & 4C +67.05 & 0.523 & Q     & 11.67 & 8.2   & H09   & 1.07  & F11   & 12.66$^{\ast}$  & 4.29  & X16  \\
    0234+285 & 4C +28.07 & 1.207 & Q     & 12.3  & 16    & S10   & 2.00  & F11   & 13.59  & 11.3  & F02    \\
    0235+164 & PKS 0235+164 & 0.94  & B     & 2     & 24    & H09   & 2.17  & F11   & 13.24  & 10.75 & W11   \\
    0333+321 & NRAO 140 & 1.259 & Q     & 12.8  & 22    & S10   & 0.36  & F11   & 13.55  & 0.61  & W11   \\
    0336-019 & PKS 0336-01 & 0.852 & Q     & 22.4  & 17.2  & S10   & 1.50  & F11   & 13.40  & 12.77 & W11    \\
    0420-014 & PKS 0420-01 & 0.915 & Q     & 7.3   & 19.7  & S10   & 1.94  & F11   & 12.88  & 17.54 & W11   \\
    0440-00 & PKS 0440-00 & 0.844 & Q & 6.1 & 11.46 & F04 & 1.30 & F11 & 13.36$^{\ast}$ & 12.6 & F04 \\
    0458-020 & 4C -02.19 & 2.286 & Q     & 16.5  & 15.7  & S10   & 0.70  & F11   & 13.50  & 17.3  & F04    \\
    0528+134 & PKS 0528+134 & 2.07  & Q     & 19.2  & 30.9  & S10   & -0.13  & F11   & 12.53  & 0.3   & F04    \\
    0552+398 & B20552+39A & 2.363 & Q     & 0.45  & 25.2  & H09   & 0.134  & F10   & 13.02$^{\ast}$  & 1.49  & W11    \\
    0605-085 & PKS 0605-08 & 0.872 & Q     & 16.8  & 7.5   & S10   & 0.09  & F11   & 13.88  & 4.61  & W11    \\
    0642+449 & S40642+449 & 3.396 & Q     & 0.8   & 10.6  & S10   & 0.51  & F11   & 13.50$^{\ast}$  & 1.67  & X16   \\
    0716+714 & S5 0716+71 & 0.31  & B     & 10.1  & 10.8  & S10   & 0.58  & F11   & 14.96  & 19.5  & W11     \\
    0735+178 & PKS 0735+17 & 0.424 & B & 5.84 & 3.17 & F04 & -0.008 & F10 & 13.02$^{\ast}$ & 36.0 & F04 \\
    0736+017 & PKS0736+01 & 0.191 & Q     & 14.4  & 8.5   & S10   & 2.16  & F11   & 14.43  & 2.2   & W11    \\
    0754+100 & PKS0754+100 & 0.266 & B     & 14.4  & 5.5   & S10   & 1.46  & F11   & 14.05  & 17.88 & W11   \\
    0804+499 & OJ 508 & 1.436 & Q     & 1.8   & 35.2  & S10   & 0.543  & F10   & 13.28  & 7.35  & W11    \\
    0814+425 & S4 0814+42 & 0.245 & B     & 1.7   & 4.6   & S10   & 0.637  & F10   & 13.52  & 9.27  & W92    \\
    0827+243 & B20827+24 & 0.941 & Q     & 22    & 13    & S10   & 1.50  & F11   & 13.50  & 2.66  & W11   \\
    0836+710 & 4C +71.07 & 2.218 & Q     & 25.4  & 16.1  & S10   & 1.27  & F11   & 14.44  & 0.77  & W11    \\
    0851+202 & PKS0851+202 & 0.306 & B     & 5.2   & 16.8  & S10   & 3.40  & F11   & 14.21  & 27.8  & W11    \\
    0923+392 & 4C +39.25 & 0.695 & Q     & 0.6   & 4.3   & S10   & 1.38  & F11   & 12.90$^{\ast}$  & 0.66  & W11   \\
    0945+408 & 4C +40.24 & 1.249 & Q     & 18.6  & 6.3   & S10   & 0.64  & W92   & 13.86  & 0.74  & W11   \\
    0954+658 & S4 0954+658 & 0.367 & B & 5.7 & 6.62 & F04 & 2.05 & F11 & 13.02$^{\ast}$ & 33.7 & F04 \\
    1055+018 & 4C +01.28 & 0.888 & B     & 8.1   & 12.1  & S10   & 0.10  & F11   & 13.79  & 16.6  & W11   \\
    1156+295 & 4C +29.45 & 0.7245 & Q     & 24.9  & 28.2  & S10   & 0.90  & F11   & 13.04  & 31.27 & W11   \\
    1219+285 & ON 231 & 0.103 & B & 2.0 & 1.56 & F04 & 0.188 & F10 & 13.32$^{\ast}$ & 20.0 & F04 \\
    1222+216 & PKS 1222+21 & 0.432 & Q     & 21    & 5.2   & S10   & 0.41  & F11   & 14.53  &       &       \\
    1226+023 & 3C 273 & 0.158 & Q     & 13.4  & 16.8  & S10   & 0.66  & F11   & 15.12  & 1.27  & W11   \\
    1253-055 & 3C 279 & 0.5362 & Q     & 20.6  & 23.8  & S10   & 0.72  & F11   & 12.69  & 44    & F02   \\
    1308+326 & B2 1308+32 & 0.997 & Q     & 20.9  & 15.3  & S10   & 1.64  & F11   & 13.22  & 19.8  & W11  \\
    1334-127 & PKS 1335-127 & 0.539 & Q     & 10.3  & 83    & S10   & 1.110  & F10   & 13.25  & 16.1  & F02    \\
    1413+135 & PKS 1413+135 & 0.247 & Q     & 1.8   & 12.1  & S10   & 0.52  & F11   & 12.57  &       &         \\
    1502+106 & PKS 1502+106 & 1.839 & Q     & 14.8  & 11.9  & S10   & 1.80  & W92   & 13.34  & 10.07 & W11   \\
    1510-089 & PKS 1510-08 & 0.361 & Q     & 20.2  & 16.5  & S10   & 1.35  & F11   & 13.97  & 8.6   & W11     \\
    1538+149 & 4C +14.60 & 0.605 & B     & 8.7   & 4.3   & S10   & 1.19  & F11   & 13.97  & 32.9  & W11    \\
    1606+106 & 4C +10.45 & 1.226 & Q     & 17.9  & 24.8  & S10   & 0.307  & F10   & 13.39  & 2.1   & F04    \\
    1611+343 & OS 319 & 1.397 & Q     & 5.7   & 13.6  & S10   & 1.40  & F11   & 13.44  & 5.3   & W11     \\
    1633+382 & 4C +38.41 & 1.814 & Q     & 29.5  & 21.3  & S10   & 1.90  & F11   & 13.21  & 3.3   & W11    \\
    1637+574 & S4 1637+574 & 0.751 & Q     & 10.6  & 13.9  & S10   & 1.44  & F11   & 14.22  & 1.4   & W11    \\
    1641+399 & 3C 345 & 0.593 & Q     & 19.3  & 7.7   & S10   & 1.33  & F11   & 13.46  & 5.78  & W11   \\
    1730-130 & NRAO 530 & 0.902 & Q     & 35.7  & 10.6  & S10   & 1.80  & F11   & 12.62  &       &       \\
    1749+096 & OT 081 & 0.322 & B     & 6.8   & 11.9  & S10   & 3.13  & F11   & 12.99  & 3.39  & W11    \\
    1803+784 & S5 1803+784 & 0.684 & B     & 9     & 12.1  & S10   & 0.28  & F11   & 13.90  & 35.2  & F02    \\
    1807+698 & 3C 371 & 0.046 & B     & 0.1   & 1.1   & S10   & 0.53  & F11   & 14.60  &       &       \\
    1823+568 & 4C +56.27 & 0.664 & B     & 9.4   & 6.3   & S10   & 0.72  & F11   & 13.25  & 0.3   & W11    \\
    1928+738 & 4C 73.18 & 0.302 & Q     & 7.2   & 1.9   & H09   & -0.28  & F11   & 13.21$^{\ast}$  & 1.56  & W11    \\
    2007+777 & S5 2007+777 & 0.342 & B & 2.33 & 5.13 & F04 & 1.27 & F11 & 12.84$^{\ast}$ & 15.1 & F04 \\
    2121+053 & PKS 2121+053 & 1.941 & Q     & 8.4   & 15.2  & S10   & 2.40  & F11   & 13.40  &       &         \\
    2134+004 & PKS 2134+004 & 1.932 & Q     & 2     & 16    & S10   & -0.68  & F11   & 12.74$^{\ast}$  & 1.01  & W11     \\
    2136+141 & PKS 2136+141 & 2.427 & Q     & 3     & 8.2   & S10   & -0.077  & F10   & 13.01$^{\ast}$  & 2.11  & X16    \\
    2145+067 & 4C +06.69 & 0.99  & Q     & 2.2   & 15.5  & S10   & 1.480  & F10   & 12.96$^{\ast}$  & 0.64  & W11   \\
    2200+420 & BL LAC & 0.069 & B     & 5     & 7.2   & S10   & 2.29  & F11   & 15.10  & 13.47 & W11     \\
    2201+315 & 4C 31.63 & 0.295 & Q     & 7.9   & 6.6   & S10   & 0.92  & F11   & 14.43  & 0.51  & W11     \\
    2223-052 & 3C 446 & 1.404 & Q     & 14.6  & 15.9  & S10   & 1.50  & F11   & 13.24  & 19.23 & W11    \\
    2230+114 & 4C -11.69 & 1.037 & Q     & 15.4  & 15.5  & S10   & 1.40  & F11   & 13.65  & 0.8   & W11    \\
    2251+158  & 3C 454.3 & 0.859 & Q     & 14.86 & 33.2  & S10   & 1.13  & F11   & 13.54  & 3.43  & W11    \\
\hline
\end{tabular}
\tablenotes{
Here, F02: Fan (2002); F04: Fan {\em et al.} (2004); F10: Fan {\em et al.} (2010); F11: Fan {\em et al.} (2011); H09: Hovatta {\em et al.} (2009); S10: Savolainen {\em et al.} (2010); W92: Wills {\em et al.} (1992); W11: Wills {\em et al.} (2011); X16: Xiao {\em et al.} (2016).
}
\end{scriptsize}
\end{table*}

\begin{table*}
\centering
\begin{scriptsize}
%\tabularfont
\caption{Calculation Results}\label{tbl:result}
\begin{tabular}{lrccccccccccc}
\topline
%\rotate
\textbf{Name}&\textbf{$\Gamma$}&\textbf{$\theta$}&\textbf{$\beta$}&\textbf{$\log f_{2+\alpha}$}&\textbf{$\log f_{3+\alpha}$}&\textbf{$\log R_{T(2+\alpha)}$}&\textbf{$\log R_{T(3+\alpha)}$}&\textbf{$\log \nu_{\rm p}^{\rm in}$}&\textbf{$\log P^{\rm in}_{2+\alpha}$}&\textbf{$\log P^{\rm in}_{3+\alpha}$}&\textbf{$\eta_{2+\alpha}$}&\textbf{$\eta_{3+\alpha}$}\\
%\midline
    (1) & (2) & (3) & (4) & (5) & (6) & (7) & (8) & (9) & (10) & (11) & (12) & (13)\\
    \hline
    0003-066 & 2.73  & 3.99  & 0.930  & -0.91  & -1.62  & -1.48  & -2.62  & 12.20  & -2.39  & -3.06  & 0.04  & 0.04  \\
    0016+731 & 6.84  & 7.29  & 0.989  & -1.19  & -2.09  & -2.56  & -4.29  & 11.87  &       &       &       &  \\
    0059+581 & 11.45  & 3.96  & 0.996  & -0.85  & -2.00  & -2.66  & -4.87  & 12.05  &       &       &       &  \\
    0106+013 & 28.42  & 2.94  & 0.999  & -1.81  & -3.07  & -4.42  & -7.13  & 12.76  & -2.96  & -4.21  & 0.08  & 0.08  \\
    0133+476 & 14.40  & 2.53  & 0.998  & -1.71  & -3.03  & -3.73  & -6.20  & 11.64  & -2.78  & -4.08  & 0.10  & 0.10  \\
    0202+149 & 8.90  & 2.77  & 0.994  & -2.08  & -3.26  & -3.68  & -5.81  & 11.07  &       &       &       &  \\
    0212+735 & 7.70  & 6.81  & 0.992  & -2.00  & -2.92  & -3.47  & -5.28  & 12.95  & -3.04  & -3.96  & 0.10  & 0.10  \\
    0219+428 & 56.95 & 7.55 & 1.000 & -0.44   & -0.74  & -3.65  & -5.70  & 14.63  & -1.19  & -1.43  & 0.32  & 0.32  \\
    0224+671 & 12.47  & 6.58  & 0.997  & -0.76  & -1.67  & -2.65  & -4.66  & 11.93  & -2.19  & -3.04  & 0.05  & 0.05  \\
    0234+285 & 12.76  & 3.46  & 0.997  & -0.41  & -1.61  & -2.32  & -4.63  & 12.73  & -1.50  & -2.57  & 0.13  & 0.13  \\
    0235+164 & 12.10  & 0.40  & 0.997  & -0.59  & -1.97  & -2.46  & -4.92  & 12.14  & -1.66  & -2.94  & 0.12  & 0.12  \\
    0333+321 & 14.75  & 2.27  & 0.998  & -2.32  & -3.67  & -4.36  & -6.87  & 12.56  & -4.53  & -5.87  & 0.01  & 0.01  \\
    0336-019 & 23.22  & 3.22  & 0.999  & -0.97  & -2.21  & -3.40  & -6.00  & 12.43  & -1.91  & -3.10  & 0.15  & 0.15  \\
    0420-014 & 11.23  & 1.90  & 0.996  & -0.65  & -1.94  & -2.45  & -4.79  & 11.87  & -1.49  & -2.70  & 0.21  & 0.21  \\
    0440-00 & 7.40  & 4.17  & 0.991  & -0.82  & -1.88  & -2.26  & -4.18  & 12.57  & -1.78  & -2.78  & 0.14  & 0.14  \\
    0458-020 & 16.55  & 3.65  & 0.998  & -1.69  & -2.89  & -3.83  & -6.24  & 12.82  & -2.46  & -3.64  & 0.21  & 0.21  \\
    0528+134 & 21.43  & 1.66  & 0.999  & -3.11  & -4.60  & -5.47  & -8.29  & 11.53  & -5.61  & -7.10  & 0.00  & 0.00  \\
    0552+398 & 12.62  & 0.08  & 0.997  & -2.67  & -4.07  & -4.57  & -7.07  & 12.15  & -4.48  & -5.88  & 0.02  & 0.02  \\
    0605-085 & 22.63  & 5.69  & 0.999  & -1.66  & -2.54  & -4.07  & -6.30  & 13.28  & -2.96  & -3.83  & 0.05  & 0.05  \\
    0642+449 & 5.38  & 0.82  & 0.983  & -1.54  & -2.57  & -2.70  & -4.46  & 13.11  & -3.32  & -4.33  & 0.02  & 0.02  \\
    0716+714 & 10.17  & 5.30  & 0.995  & -1.49  & -2.52  & -3.20  & -5.24  & 14.04  & -2.20  & -3.22  & 0.25  & 0.25  \\
    0735+178 & 7.12 & 15.15 & 0.990  & -1.01   & -1.51   & -2.41   & -3.77   & 12.67  & -1.37  & -1.85  & 0.91  & 0.91  \\
    0736+017 & 16.51  & 5.90  & 0.998  & 0.30  & -0.63  & -1.83  & -3.98  & 13.58  & -1.83  & -2.38  & 0.02  & 0.02  \\
    0754+100 & 21.69  & 6.94  & 0.999  & -0.02  & -0.76  & -2.39  & -4.47  & 13.42  & -1.06  & -1.58  & 0.22  & 0.22  \\
    0804+499 & 17.66  & 0.17  & 0.998  & -2.55  & -4.10  & -4.74  & -7.54  & 12.12  & -3.68  & -5.23  & 0.08  & 0.08  \\
    0814+425 & 2.72  & 8.39  & 0.930  & -0.69  & -1.35  & -1.26  & -2.36  & 12.95  & -1.78  & -2.38  & 0.11  & 0.11  \\
    0827+243 & 25.15  & 3.86  & 0.999  & -0.73  & -1.84  & -3.23  & -5.74  & 12.67  & -2.38  & -3.42  & 0.03  & 0.03  \\
    0836+710 & 28.12  & 3.22  & 0.999  & -1.14  & -2.35  & -3.74  & -6.40  & 13.74  & -3.29  & -4.46  & 0.01  & 0.01  \\
    0851+202 & 9.23  & 1.93  & 0.994  & 0.95  & -0.28  & -0.68  & -2.87  & 13.10  & -0.60  & -1.02  & 0.39  & 0.39  \\
    0923+392 & 2.31  & 3.85  & 0.901  & 0.11  & -0.52  & -0.31  & -1.31  & 12.49  & -2.42  & -2.81  & 0.01  & 0.01  \\
    0945+408 & 30.69  & 5.52  & 0.999  & -0.96  & -1.76  & -3.63  & -5.92  & 13.41  & -3.12  & -3.88  & 0.01  & 0.01  \\
    0954+658 & 5.84 & 8.61  & 0.985  & 0.41   & -0.41  & -0.82  & -2.41  & 12.34  & -0.62  & -1.03  & 0.51  & 0.51  \\
    1055+018 & 8.80  & 4.39  & 0.994  & -2.07  & -3.15  & -3.65  & -5.68  & 12.98  & -2.82  & -3.90  & 0.21  & 0.21  \\
    1156+295 & 25.11  & 2.02  & 0.999  & -2.00  & -3.45  & -4.50  & -7.35  & 11.83  & -2.51  & -3.95  & 0.46  & 0.46  \\
    1219+285 & 2.38 & 36.36 & 0.908  & -0.21   & -0.39   & -0.66   & -1.22   & 13.17  & -0.96  & -1.09  & 0.40  & 0.40  \\
    1222+216 & 45.10  & 5.14  & 1.000  & -1.02  & -1.74  & -4.03  & -6.40  & 13.97  &       &       &       &  \\
    1226+023 & 13.77  & 3.33  & 0.997  & -1.79  & -3.02  & -3.77  & -6.13  & 13.96  & -3.69  & -4.91  & 0.01  & 0.01  \\
    1253-055 & 20.84  & 2.38  & 0.999  & -2.03  & -3.41  & -4.37  & -7.07  & 11.50  & -2.39  & -3.76  & 0.80  & 0.80  \\
    1308+326 & 21.96  & 3.57  & 0.999  & -0.73  & -1.91  & -3.11  & -5.64  & 12.34  & -1.51  & -2.62  & 0.25  & 0.25  \\
    1334-127 & 42.15  & 0.17  & 1.000  & -2.73  & -4.65  & -5.68  & -9.22  & 11.51  & -3.52  & -5.44  & 0.19  & 0.19  \\
    1413+135 & 6.23  & 1.39  & 0.987  & -1.65  & -2.73  & -2.93  & -4.81  & 11.58  &       &       &       &  \\
    1502+106 & 15.20  & 4.70  & 0.998  & -0.35  & -1.43  & -2.41  & -4.67  & 12.72  & -1.51  & -2.44  & 0.11  & 0.11  \\
    1510-089 & 20.65  & 3.40  & 0.999  & -1.08  & -2.30  & -3.41  & -5.95  & 12.88  & -2.18  & -3.37  & 0.09  & 0.09  \\
    1538+149 & 11.07  & 10.58  & 0.996  & -0.08  & -0.71  & -1.86  & -3.54  & 13.54  & -0.82  & -1.26  & 0.50  & 0.50  \\
    1606+106 & 18.88  & 2.19  & 0.999  & -2.48  & -3.88  & -4.73  & -7.40  & 12.35  & -4.15  & -5.55  & 0.02  & 0.02  \\
    1611+343 & 8.03  & 3.01  & 0.992  & -0.87  & -2.00  & -2.38  & -4.41  & 12.69  & -2.20  & -3.28  & 0.06  & 0.06  \\
    1633+382 & 31.10  & 2.55  & 0.999  & -0.76  & -2.09  & -3.44  & -6.26  & 12.33  & -2.31  & -3.57  & 0.03  & 0.03  \\
    1637+574 & 11.03  & 3.98  & 0.996  & -0.85  & -1.99  & -2.63  & -4.82  & 13.32  & -2.76  & -3.85  & 0.01  & 0.01  \\
    1641+399 & 28.10  & 5.12  & 0.999  & -0.44  & -1.33  & -3.04  & -5.37  & 12.78  & -1.81  & -2.58  & 0.06  & 0.06  \\
    1730-130 & 65.46  & 2.95  & 1.000  & -0.25  & -1.28  & -3.58  & -6.42  & 11.87  &       &       &       &  \\
    1749+096 & 7.93  & 4.16  & 0.992  & 0.98  & -0.10  & -0.52  & -2.49  & 12.04  & -1.51  & -1.82  & 0.04  & 0.04  \\
    1803+784 & 9.44  & 4.55  & 0.994  & -1.89  & -2.97  & -3.53  & -5.59  & 13.05  & -2.33  & -3.40  & 0.58  & 0.58  \\
    1807+698 & 1.01  & 42.27  & 0.134  & 0.22  & 0.22  & 0.52  & 0.50  & 14.58  &       &       &       &  \\
    1823+568 & 10.24  & 8.42  & 0.995  & -0.88  & -1.68  & -2.60  & -4.41  & 12.67  & -3.44  & -4.20  & 0.00  & 0.00  \\
    1928+738 & 14.86  & 14.81  & 0.998  & -0.84  & -1.12  & -2.88  & -4.33  & 13.05  & -2.40  & -2.65  & 0.03  & 0.03  \\
    2007+777 & 3.19 & 8.62   & 0.950  & -0.15   & -0.86  & -0.86  & -2.07  & 12.26  & -1.20  & -1.73  & 0.18  & 0.18  \\
    2121+053 & 9.95  & 3.20  & 0.995  & 0.04  & -1.15  & -1.66  & -3.84  & 12.69  &       &       &       &  \\
    2134+004 & 8.16  & 0.88  & 0.992  & -3.09  & -4.29  & -4.61  & -6.73  & 12.00  & -4.97  & -6.17  & 0.01  & 0.01  \\
    2136+141 & 4.71  & 4.56  & 0.977  & -1.90  & -2.82  & -2.95  & -4.54  & 12.63  & -3.53  & -4.44  & 0.02  & 0.02  \\
    2145+067 & 7.94  & 1.03  & 0.992  & -0.90  & -2.09  & -2.40  & -4.49  & 12.07  & -3.14  & -4.29  & 0.01  & 0.01  \\
    2200+420 & 5.41  & 7.51  & 0.983  & 0.58  & -0.28  & -0.59  & -2.18  & 14.27  & -0.97  & -1.33  & 0.16  & 0.16  \\
    2201+315 & 8.10  & 8.56  & 0.992  & -0.72  & -1.54  & -2.24  & -3.96  & 13.73  & -3.08  & -3.84  & 0.01  & 0.01  \\
    2223-052 & 14.68  & 3.59  & 0.998  & -0.90  & -2.10  & -2.94  & -5.30  & 12.42  & -1.67  & -2.82  & 0.24  & 0.24  \\
    2230+114 & 15.43  & 3.70  & 0.998  & -0.98  & -2.17  & -3.06  & -5.44  & 12.76  & -3.12  & -4.27  & 0.01  & 0.01  \\
    2251+158  & 19.94  & 1.29  & 0.999  & -1.91  & -3.43  & -4.21  & -7.03  & 12.29  & -3.38  & -4.90  & 0.04  & 0.04  \\
\hline
\end{tabular}
\tablenotes{
Col. (1) gives sources name,
Col. (2) Lorentz factor,
Col. (3) viewing angle,
Col. (4) jet speed in units of speed of light,
Col. (5) ratio $f$ for the case of $q=2+\alpha$,
Col. (6) ratio $f$ for $q=3+\alpha$,
Col. (7) core-dominance parameter at 90 degree for $q=2+\alpha$,
Col. (8) core-dominance parameter at 90 degree for $q=3+\alpha$,
Col. (9) intrinsic peak frequency in units of Hz,
Col. (10) intrinsic polarization for $q=2+\alpha$,
Col. (11) intrinsic polarization for $q=3+\alpha$,
Col. (12) ratio $\eta$ for $q=2+\alpha$,
Col. (13) ratio $\eta$ for $q=3+\alpha$.
}
\end{scriptsize}
\end{table*}

\clearpage

\begin{table*}
\small
\caption{Averaged Values}
\label{tbl:Averaged}
\begin{tabular}{|c|cc|cc|cc|}
\hline
\multirow{2}{1.8cm}{Parameter} & \multicolumn{2}{c|}{All} & \multicolumn{2}{c|}{BL Lacs} & \multicolumn{2}{c|}{FSRQs}\\
\cline{2-7}
& Range & Average & Range & Average & Range & Average\\
\hline
$\delta$ & 1.1--83 & 14.23$\pm$11.64 & 1.1--24 & 7.79$\pm$5.90 & 1.9--83 & 16.69$\pm$12.38 \\
$\beta_{app}$ & 0.1--35.7 & 11.30$\pm$8.11 & 0.1--14.9 & 6.23$\pm$4.37 & 0.45--35.7 & 13.25$\pm$8.39\\
$\log R$ & $-$0.68--3.4 & 1.02$\pm$0.92 & -0.01--3.40 & 1.15$\pm$1.06 & $-$0.68--2.4 & 0.98$\pm$0.71\\
$\log \nu _{\rm p}^{\rm s}$ & 12.10--15.12 & 13.46$\pm$0.67 & 12.78--15.10 & 13.74$\pm$0.75 & 12.10--15.12 & 13.35$\pm$0.61\\
$\log P^{\rm ob}$ & $-$2.52--$-$0.36 & $-$1.28$\pm$0.60 & $-$2.52--$-$0.44 & $-$0.90$\pm$0.52 & $-$2.52--$-$0.36 & $-$1.44$\pm$0.56\\
$\Gamma$ & 1.01--65.47 & 15.75$\pm$12.32 & 1.01--56.95 & 10.45$\pm$12.57 & 2.31--65.46 & 17.79$\pm$11.72\\
$\theta$ & $0.08^{\circ}$--$42.27^{\circ}$ & $5.46^{\circ}\pm6.82^{\circ}$ & $0.40^{\circ}$--$42.27^{\circ}$ & $10.28^{\circ}\pm11.12^{\circ}$ & $0.08^{\circ}$--$14.81^{\circ}$ & $3.61^{\circ}\pm2.52^{\circ}$\\
$\beta$ & 0.13--1.00 & 0.98$\pm$0.11 & 0.13--1.00 & 0.93$\pm$0.20 & 0.90--1.00 & 0.99$\pm$0.01\\
$\log f (q=2+\alpha)$ & $-$3.11--0.98 & $-$1.06$\pm$0.93 & $-$2.07--0.98 & $-$0.40$\pm$0.88 & $-$3.11--0.30 & $-$1.31$\pm$0.84\\
$\log f (q=3+\alpha)$ & $-$4.65--0.22 & $-$2.09$\pm$1.15 & $-$3.15--0.22 & $-$1.17$\pm$0.99 & $-$4.65--$-$0.52 & $-$2.45$\pm$1.01\\
$\log R_T (q=2+\alpha)$ & $-$5.68--0.52 & $-2.91\pm1.30$ & $-$3.65--0.52 & $-1.78\pm1.26$ & $-$5.68--$-$0.31 & $-3.34\pm1.03$\\
$\log R_T (q=3+\alpha)$ & $-$9.22--0.50 & $-5.02\pm1.79$ & $-$5.70--0.50 & $-3.39\pm1.72$ & $-$9.22--$-$1.31 & $-5.65\pm1.39$\\
$\log \nu _{\rm p}^{\rm in}$ & 11.07--14.62 & 12.69$\pm$0.77 & 12.04--14.62 & 13.11$\pm$0.83 & 11.07--13.97 & 12.52$\pm$0.69\\
$\log P^{\rm in}(q=2+\alpha)$ & $-$5.61--$-$0.60 & $-$2.46$\pm$1.10 & $-$3.44--$-$0.60 & $-$1.58$\pm$0.81 & $-$5.61--$-$1.49 & $-$2.84$\pm$0.99\\
$\log P^{\rm in}(q=3+\alpha)$ & $-$7.10--$-$1.02 & $-$3.43$\pm$1.38 & $-$4.20--$-$1.02 & $-$2.19$\pm$1.06 & $-$7.10--$-$2.38 & $-$3.95$\pm$1.15\\
$\eta (q=2+\alpha)$ & 0.003--0.91 & 0.15$\pm$0.20 & 0.003--0.91 & 0.29$\pm$0.24 & 0.003--0.80 & 0.10$\pm$0.15\\
$\eta (q=3+\alpha)$ & 0.003--0.91 & 0.15$\pm$0.20 & 0.003--0.91 & 0.29$\pm$0.24 & 0.003--0.80 & 0.10$\pm$0.15\\
\hline
\end{tabular}
%\tablecomments{****}
\end{table*}

\begin{table*}
\small
\caption{Correlations between Ratio, $f$, and Other Parameters for Blazars.}
\label{tbl:Correlation}
\begin{tabular}{|c|c|c|c|c|c|c|c|c|c|}
\hline
  $y$ & $x$ & Jet Case & Sample & N & $A \pm \sigma$ & $B \pm \sigma$ & $r$ & Probability & Figure\\
\hline
  (1) & (2) & (3) & (4) & (5) & (6) & (7) & (8) & (9) & (10)\\
\hline
  \multirow{6}{1cm}{$\log f$} & \multirow{6}{1cm}{$\log \nu^{\rm s}_{\rm p}$} & \multirow{3}{1.25cm}{$q=2+\alpha$}& all & 65 &  $0.36 \pm 0.17$ & $-5.93 \pm 2.28$ & 0.26 & $3.63\%$ & \multirow{6}{0.9cm}{Fig. \ref{Lin-2016-logf-lognu}} \\
  &&& BL Lacs & 18 &  $0.04 \pm 0.29$ & $-1.00 \pm 4.04$ & 0.04 & $88.54\%$ &  \\
  &&& FSRQs & 47 &  $0.32 \pm 0.20$ & $-5.53 \pm 2.64$ & 0.23 & $11.62\%$ &  \\
  \cline{3-9}
  && \multirow{3}{1.25cm}{$q=3+\alpha$} & all & 65 &  $0.48 \pm 0.21$ & $-8.57 \pm 2.79$ & 0.28 & $2.35\%$ & \\
  &&& BL Lacs & 18 &  $0.09 \pm 0.33$ & $-2.42 \pm 4.54$ & 0.07 & $78.59\%$ &  \\
  &&& FSRQs & 47 &  $0.38 \pm 0.24$ & $-7.49 \pm 3.18$ & 0.23 & $11.97\%$ &  \\
  \hline
  \multirow{6}{1cm}{$\log f$} & \multirow{6}{1cm}{$\log \nu_{\rm p}^{\rm in}$} & \multirow{3}{1.25cm}{$q=2+\alpha$}& all & 65 &  $0.45 \pm 0.14$ & $-6.76 \pm 1.81$ & 0.37 & $2.50\times10^{-3}$ & \multirow{6}{0.9cm}{Fig. \ref{Lin-2016-logf-lognuin}} \\
  &&& BL Lacs & 18 &  $0.03 \pm 0.27$ & $-0.85 \pm 3.50$ & 0.03 & $90.01\%$ &  \\
  &&& FSRQs & 47 &  $0.45 \pm 0.17$ & $-6.89 \pm 2.12$ & 0.37 & $1.13\%$ &  \\
  \cline{3-9}
  && \multirow{3}{1.25cm}{$q=3+\alpha$} & all & 65 &  $0.70 \pm 0.17$ & $-10.94 \pm 2.12$ & 0.47 & $9.40 \times 10^{-5}$ & \\
  &&& BL Lacs & 18 &  $0.24 \pm 0.29$ & $-4.30 \pm 3.87$ & 0.20 & $42.96\%$ &  \\
  &&& FSRQs & 47 &  $0.64 \pm 0.20$ & $-10.40 \pm 2.48$ & 0.43 & $2.40\times10^{-3}$ &  \\
  \hline
  \multirow{6}{1cm}{$\log {\rm P}^{\rm ob}$} & \multirow{6}{1cm}{$\log f$} & \multirow{3}{1.25cm}{$q=2+\alpha$} & all & 57 & $0.14 \pm 0.08$ & $-1.13 \pm 0.12$ & 0.22 & $10.02\%$ & \multirow{6}{0.9cm}{Fig. \ref{Lin-2016-logP-logf}} \\
  &&& BL Lacs & 17 &  $0.01 \pm 0.15$ & $-0.89 \pm 1.15$ & 0.01 & $97.35\%$ &  \\
  &&& FSRQs & 40 &  $0.04 \pm 0.11$ & $-1.39 \pm 0.17$ & 0.06 & $73.01\%$ &  \\
  \cline{3-9}
  && \multirow{3}{1.25cm}{$q=3+\alpha$} & all & 57 &  $0.10 \pm 0.07$ & $-1.06 \pm 0.17$ & 0.20 & $14.26\%$ & \\
  &&& BL Lacs & 17 &  $0.02 \pm 0.14$ & $-0.87 \pm 0.22$ & 0.03 & $90.34\%$ &  \\
  &&& FSRQs & 40 &  $-0.02 \pm 0.09$ & $-1.48 \pm 0.24$ & $-$0.03 & $83.97\%$ &  \\
  \hline
  \multirow{6}{1cm}{$\log {\rm P}^{\rm in}$} & \multirow{6}{1cm}{$\log f$} & \multirow{3}{1.25cm}{$q=2+\alpha$} & all & 57 & $0.96 \pm 0.09$ & $-1.42 \pm 0.12$ & 0.84 & $6.21 \times 10^{-16}$ & \multirow{6}{0.9cm}{Fig. \ref{Lin-2016-logPin-logf}} \\
  &&& BL Lacs & 17 &  $0.67 \pm 0.16$ & $-1.29 \pm 0.15$ & 0.74 & $7.55\times10^{-4}$ &  \\
  &&& FSRQs & 40 &  $0.95 \pm 0.11$ & $-1.54 \pm 0.18$ & 0.82 & $1.48 \times 10^{-10}$ &  \\
  \cline{3-9}
  && \multirow{3}{1.25cm}{$q=3+\alpha$} & all & 57 &  $1.07 \pm 0.07$ & $-1.13 \pm 0.17$ & 0.91 & $4.59 \times 10^{-22}$ &  \\
  &&& BL Lacs & 17 &  $0.95 \pm 0.14$ & $-1.00 \pm 0.22$ & 0.87 & $8.00\times10^{-6}$ &  \\
  &&& FSRQs & 40 &  $0.97 \pm 0.09$ & $-1.50 \pm 0.24$ & 0.87 & $1.86 \times 10^{-13} $&  \\
  \hline
  \multirow{6}{1cm}{$\log f$} & \multirow{6}{1cm}{$\log R$} & \multirow{3}{1.25cm}{$q=2+\alpha$} & all & 65 & $0.83 \pm 0.10$ & $-1.90 \pm 0.13$ & 0.72 & $1.57 \times 10^{-11}$ & \multirow{6}{0.9cm}{Fig. \ref{Lin-2016-logf-logR}} \\
  &&& BL Lacs & 18 &  $0.64 \pm 0.14$ & $-1.14 \pm 0.21$ & 0.77 & $2.19\times10^{-4}$ &  \\
  &&& FSRQs & 47 &  $0.91 \pm 0.11$ & $-2.19 \pm 0.14$ & 0.77 & $3.13 \times 10^{-10}$ &  \\
  \cline{3-9}
  && \multirow{3}{1.25cm}{$q=3+\alpha$} & all & 65 &  $0.74 \pm 0.15$ & $-2.85 \pm 0.20$ & 0.52 & $8.00 \times 10^{-6}$ &  \\
  &&& BL Lacs & 18 &  $0.45 \pm 0.21$ & $-1.69 \pm 0.32$ & 0.48 & $4.29\%$ &  \\
  &&& FSRQs & 47 &  $0.86 \pm 0.17$ & $-3.29 \pm 0.20$ & 0.60 & $7.00 \times 10^{-6} $&  \\
\hline
\end{tabular}
\tablenotes{
Col. (1) gives dependent parameter,
Col. (2) independent parameter,
Col. (3) jet case, $q = 2+\alpha$ stand for the stationary jet and $q = 3+\alpha$ for the jet with distinct ¡°blobs¡±,
Col. (4) sample,
Col. (5) number of the sample,
Col. (6) slope,
Col. (7) intercept,
Col. (8) correlation coefficient,
Col. (9) chance probability,
Col. (10) the corresponding figure.
}
\end{table*}

%%Appendix


\begin{thebibliography}{99}
\bibitem{Abdo10} Abdo, A. A., Ackermann, M., Agudo, I., et al., 2010, {\em ApJ}, {\bf 716}, 30.
\bibitem{Acker15} Ackermann, M.; Ajello, M.; Atwood, W. B., et al., 2015, {\em ApJ}, {\bf 810}, 14.
\bibitem{Anton85} Antonucci, R. R. J. \& Ulvestad, J. S. 1985, {\em ApJ}, {\bf 294}, 158.

\bibitem{Bland79} Blandford, R. D. \& K\"onigl, A. 1979, {\em ApJ}, {\bf 232}, 34.
%\bibitem[Cao et al. (1999)]{Xinwu99} Cao X. W., Jiang D. R., et al., 1999, MNRAS, 307, 802

\bibitem{Fan97} Fan, J. H., Cheng, K. S., Zhang, L., Liu, C. H., 1997, {\em A\&A}, {\bf 327}, 947.
\bibitem{Fan02} Fan, J. H. 2002, {\em PASJ}, {\bf 54}, L55.
\bibitem{Fan03a} Fan, J. H., 2003, {\em ApJ}, {\bf 585}, L23.
\bibitem{Fan03b} Fan, J. H. \& Lin, R. G., 2003, {\em ChPhy}, {\bf 12}, 332.
\bibitem{Fan04} Fan, J. H., Wang, Y. J., Yang, J. H., Su, C. Y., 2004, {\em ChJAA}, {\bf 4}, 533.
\bibitem{Fan05} Fan, J. H. 2005, {\em A\&A}, {\bf 436}, 799.
\bibitem{Fan06} Fan, J. H., Wang, Y. X., Hua, T. X., et al., 2006, {\em ChJAS}, {\bf 6b}, 349.
%\bibitem[Fan et al. (2006)]{Fan06} Fan, J. H., Hua, T. X., Yuan, Y. H., et al., 2006, PASJ, 58, 945
\bibitem{Fan09} Fan, J. H., Huang, Y., He, T. M., et al., 2009, {\em PASJ}, {\bf 61}, 639.
\bibitem{Fan10} Fan, J. H., Yang, J. H., Tao, J., Huang, Y., Liu, Y., 2010, {\em PASJ}, {\bf 62}, 211.
\bibitem{Fan11} Fan, J. H., Yang, J. H., Pan, J., Hua, T. X., 2011, {\em RAA}, {\bf 11}, 1413.
\bibitem{Fan14} Fan, J. H., Bastieri, D., Yang, J. H., et al., 2014, {\em RAA}, {\bf 14}, 1135.
\bibitem{Fan15} Fan, J. H., Yang, J. H., Liu, Y., Cai, W., Lin, C., 2015, {\em IJMPA}, Vol. 30, No. 32, 1530023.
\bibitem{Fan16} Fan, J. H., Yang, J. H., Liu, Y., et al., 2016, {\em ApJS}, {\bf 226}, 20.

\bibitem{Gupta09} Gupta, Alok C., Srivastava, A. K., \& Wiita, Paul J., 2009, {\em ApJ}, {\bf 690}, 216
\bibitem{Gupta16} Gupta, Alok C., Agarwal, A., Bhagwan, J., et al., 2016, {\em MNRAS}, {\bf 458}, 1127

\bibitem{Hova09} Hovatta, T., Valtaoja, E., Tornikoski, M., L\"ahteenm\"aki, A., 2009, {\em A\&A}, {\bf 496}, 527.
%\bibitem[Hutchings et al. (1988)]{Hutch88} Hutchings, J. B., Price, R., \& Gower, A. C. 1988, ApJ, 329, 122

\bibitem{Laht99} L\"ahteenm\"aki, A. \& Valtaoja, E., 1999, {\em ApJ}, {\bf 521}, 493.
\bibitem{Lin16a} Lin C. \& Fan, J. H., 2016, {\em RAA}, {\bf 16}, 103.
\bibitem{Lin17} Lin C., Fan, J. H., Xiao H. B., 2017, accepted by {\em RAA}
\bibitem{Lind85} Lind, K. R., \& Blandford, R. D. 1985, {\em ApJ}, {\bf 295}, 358.

\bibitem{Massaro15} Massaro, E., Maselli, A., Leto, C., et al., 2015, {\em Ap\&SS}, {\bf 357}, 75.

%\bibitem[Neff et al. (1989)]{Neff89} Neff, S. G., Hutchings, J. B., \& Gower, A. C. 1989, AJ, 97, 1291
%\bibitem[Neff \& Hutchings (1990)]{Neff90} Neff, S. G., \& Hutchings, J. B. 1990, AJ, 100, 1441
\bibitem{Niep08} E. Nieppola, E. Valtaoja, M. Tornikoski., et al., 2008, {\em A\&A}, {\bf 488}, 867.


\bibitem{Orr82} Orr, M. J. L. \& Browne, I. W. A., 1982, {\em MNRAS}, {\bf 200}, 1067.

\bibitem{Pado90} Padovani P. \& Urry C. M. 1990, {\em ApJ}, {\bf 356}, 75.
\bibitem{Pado91} Padovani P. \& Urry C. M. 1991, {\em ApJ}, {\bf 386}, 373.
\bibitem{Pado92} Padovani P. 1992, {\em MNRAS}, {\bf 257}, 404.
%\bibitem[Pan \& Fan (2011)]{Pan2011} Pan, J., Fan, J. H., 2011, SCPMA, 54, 2265
\bibitem{Pei16} Pei, Z. Y., Fan, J. H., Liu, Y., et al., 2016, {\em Ap\&SS}, {\bf 361}, 237

\bibitem{Samb96} Sambruna, R. M., Maraschi, L., \& Urry, C. M. 1996, {\em ApJ}, {\bf 463}, 444.
\bibitem{Savo10} Savolainen, T., Homan, D. C., Hovatta, T., et al., 2010, {\em A\&A}, {\bf 512A}, 24.

\bibitem{Urry91} Urry C. M. \& Padovani P. 1991, {\em ApJ}, {\bf 371}, 60.
\bibitem{Urry95} Urry C. M. \& Padovani P. 1995, {\em PASP}, {\bf 107}, 803.

\bibitem{Wills92} Wills, B. J., Wills, D., Breger, M., et al., 1992, {\em ApJ}, {\bf 398}, 454.
\bibitem{Wills11} Wills, B. J., Wills, D., \& Breger, M., 2011, {\em ApJS}, {\bf 194}, 19.

\bibitem{Wu02} Wu, X. B., Liu, F. K., \& Zhang, T. Z. 2002, {\em A\&A}, {\bf 389}, 742.
%\bibitem[Wu et al. (2007)]{Wu07} Wu, Z. Z., Jiang, D. R., Gu, M.F., Liu, Y., 2007, A\&A, 466, 63

%\bibitem[Xiao et al. (2015)]{Xiao15} Xiao, H. B., Pei, Z. Y., Xie, H. J., et al., 2015, ApSS, 359, 39
\bibitem{Xiao16} Xiao, H. B., et al., 2016, in prepare.

\bibitem{Yang10} Yang, J. H., Fan, J. H., \& Yang, R. S., 2010, {\em SCPMA}, {\bf 53}, 1162.
\end{thebibliography}
\end{document}